\documentclass[aip,graphicx,reprint,pop,floatfix]{revtex4-1}

\usepackage{amsmath}
\usepackage{amsfonts}
\usepackage{graphicx}
\usepackage{mathtools}
\usepackage{multirow}
\usepackage{natbib}
\newcommand{\leri}[1]{\left(#1\right)}
\newcommand{\lerisq}[1]{\left[#1\right]}

\begin{document}

%\preprint{}

\title{Kinetic Fluctuation-Dissipation Relation in collisionless plasmas: a correlated noise source} %Title of paper



\author{F. Moretti}
\email[]{fabio.moretti@cnr.it}
\affiliation{National Research Council of Italy, Institute for Plasma Science and Technology, Via G. Amendola 122/D, I-70126 Bari, Italy - SDL Tor Vergata, Rome}

\author{O. Pezzi}
\email[]{oreste.pezzi@cnr.it}
\affiliation{National Research Council of Italy, Institute for Plasma Science and Technology, Via G. Amendola 122/D, I-70126 Bari, Italy - SDL Tor Vergata, Rome}

\author{S. Benella}
\affiliation{INAF—Institute for Space Astrophysics and Planetology, Via del Fosso del Cavaliere, 100, 00133, Rome, Italy}

\author{G. Consolini}
\affiliation{INAF—Institute for Space Astrophysics and Planetology, Via del Fosso del Cavaliere, 100, 00133, Rome, Italy}

\author{A. Larosa}
\affiliation{National Research Council of Italy, Institute for Plasma Science and Technology, Via G. Amendola 122/D, I-70126 Bari, Italy - SDL Tor Vergata, Rome}

\author{F. Lepreti}
\affiliation{Department of Physics, University of Calabria, via P. Bucci, 87036 Rende (CS), Italy}
\affiliation{National Institute for Astrophysics, Scientific Directorate, Viale del Parco Mellini 84, I-00136 Rome, Italy
}

\author{L. Sorriso-Valvo}
\affiliation{National Research Council of Italy, Institute for Plasma Science and Technology, Via G. Amendola 122/D, I-70126 Bari, Italy - SDL Rende (CS)}
\affiliation{Department of Electromagnetics and Plasma Physics, School of Electrical Engineering and Computer Science, KTH Royal Institute of Technology, Teknikringen 31, SE-11428 Stockholm, Sweden}

\author{F. Valentini}
\affiliation{Department of Physics, University of Calabria, via P. Bucci, 87036 Rende (CS), Italy}

\author{V. Carbone}
\affiliation{Department of Physics, University of Calabria, via P. Bucci, 87036 Rende (CS), Italy}
\affiliation{National Institute for Astrophysics, Scientific Directorate, Viale del Parco Mellini 84, I-00136 Rome, Italy
}

\date{\today}

\begin{abstract}
We here focus on small-scale dynamics of collisionless plasmas by employing a generalized kinetic Langevin framework driven by a stochastic time-correlated forcing. Our investigation captures the steady-state Fluctuation-Dissipation Relation (FDR) where the energy injection at low frequencies is due to the time-correlated forcing and the small-scale energy conversion occurs via Landau damping, the latter --though formally reversible-- acting as an effective dissipation mechanism. We analytically extract the resulting asymptotic spectral scalings across multiple physical regimes and characterize the velocity-space cascade by decomposing the plasma distribution function with Hermite polynomials. The framework is then generalized to a multi-scale forcing scenario in which correlation rates are distributed according to a power-law distribution. This extension provides closed-form analytical expressions that may serve to interpret spacecraft in-situ observations. All analytical predictions are successfully validated through direct numerical integrations of the underlying kinetic equations.
\end{abstract}

\pacs{}% insert suggested PACS numbers in braces on next line

\maketitle %\maketitle must follow title, authors, abstract and \pacs

\section{Introduction}\label{sect:intr}

The issue of energy conversion and dissipation in nearly collisionless plasmas is a fundamental, long-standing problem in space and astrophysical plasma physics \citep{PhysRevLett.114.175002, 2018ApJ...862...32C, 2021MNRAS.505.4857P}. In these environments, such as the solar wind and the Earth's magnetosphere, viscous-like dissipation induced by collisional effects is thought to be negligible, as collisional frequencies are way smaller than typical frequencies associated with collisionless phenomena. Therefore, energy conversion at kinetic scales is mediated by nonlinear wave-particle interactions and generally occurs in the entire phase space, frequently driving non-thermal plasma distribution functions. 

A common feature of space and astrophysical plasmas is that they are manifestly turbulent \citep{2013LRSP...10....2B}. Nonlinear mechanisms transfer energy from large, energy-containing scales to small, kinetic scales at which energy conversion and, eventually, dissipation occur. Moreover, temporal and spatial coherent structures such as vortices and sheets emerge across scales as a result of intermittency. The fundamental turbulent nature of nearly-collisionless space plasmas has paved the way to investigate these systems in the framework of statistical physics of non-equilibrium complex systems \cite{Chang1999, https://doi.org/10.1029/1999JA000319, 2001SSRv...95..309C, CONSOLINI2002541, 2009PhRvL.103g5006K, 2016PhPl...23b2103P, 2024PhRvL.132h5201D}. In this context, the Fluctuation-Dissipation Relation (FDR) represents a powerful theoretical tool to describe the steady-state balance between the energy injected into the system at large scales and the internal energy conversion and dissipation mechanisms.
    
Previous theoretical models \cite{2015JPlPh..81a3004K} have made significant progress by treating the turbulent energy injection as a white-noise source within a kinetic Langevin framework. However, this approach intrinsically leaves a knowledge gap, as also evidenced by recent studies based on space plasmas observations \cite{Carbone_2022, Chiappetta_2023}. Indeed, by definition, turbulence is characterized by finite spatial and temporal correlations \cite{d573db38-d0d6-37f9-91e9-b9675eb19d41, 1982JGR....87.6011M, 1995tlan.book.....F}, and assuming instantaneous decorrelation (white noise) may oversimplify the turbulent plasma dynamics. Hence, treating turbulent forcing with time-correlated noise is important to accurately model large-scale forcing and, in turn, interpret the plasma behavior and spectral signatures~\cite{Carbone_2022, Chiappetta_2023}.

In this work, we generalize the kinetic framework based on the Langevin phenomenology introduced in Ref. \cite{2015JPlPh..81a3004K}, by considering a damped-forced collisionless electrostatic plasma model where plasma response is driven by a stochastic forcing term characterized by a finite correlation time, which mimics the property of plasma turbulence \cite{Carbone_2022}. In this framework, the continuous power injected from larger scales is counterbalanced by collisionless energy conversion mechanisms, as Landau damping. The forcing drives a stationary state where these kinetic processes effectively mediate the energy conversion and, despite their reversible nature, act as an effective dissipation, intended here as a energy transfer from electric field to particle distribution function. This balance ultimately drives the plasma toward a steady-state equilibrium governed by the FDR. We analytically derive the generalized FDR for the system response and extract its asymptotic spectral scalings across several interesting physical regimes. Moreover, to establish a direct connection between theoretical modeling and in-situ spacecraft observations, we extend our analysis to a multi-scale forcing scenario: the stochastic source is modeled as a superposition of noises with different correlation times governed by a power-law distribution \cite{Carbone_2022}. All of our analytical predictions are validated against direct numerical integrations of the underlying kinetic equations.
    
The paper is organized as follows. In Section \ref{sec2} we introduce the kinetic analogue of the Langevin equation, provide the details of the generalized FDR derivation, and explore its asymptotic limits and Hermite spectrum. Then, the numerical simulations validating the theoretical scalings are illustrated in Section \ref{sec3}, Section \ref{sec4} extends the framework to the multi-scale forcing where the correlation times follow a power-law distribution, and we summarize our conclusions in Section \ref{sec5}. 

This work is dedicated to the memory of Professor Vincenzo Carbone, mentor and friend for many of us, who sadly passed away about two years ago. His pioneering paper \cite{Carbone_2022} and later discussions with several of us undoubtedly contributed to set the theoretical foundation of the analyses shown here.

\section{Kinetic analogue of the Langevin equation}\label{sec2}
We are interested in describing the evolution of a quasi-neutral plasma, where one species is treated within a kinetic Vlasov framework and another one is a background neutralizing flow. To establish a connection with the phenomenology of the Langevin equation, a damped-forced scenario must be considered. In the model analyzed here, some of the properties of large- and ion-scale plasma turbulence are collected into a stochastic source term on the right-hand side of the Vlasov equation. This forcing represents the power injected into the system from larger scales and takes the form of a noise with a finite correlation time  \cite{Carbone_2022}. Conversely, dissipation is included in our description solely through the Landau damping mechanism. Although Landau damping is not a true collisional dissipation owing to its reversibility, it effectively mediates the transfer of energy from field to plasma particles. Nevertheless, by restricting the analysis on the wavenumbers excited through the forcing, Landau damping successfully mimics macroscopic dissipation. Indeed, it continuously drains energy from fluctuations, counterbalancing the stochastic injection and ultimately driving the plasma toward a steady-state equilibrium governed by an FDR.

Following the analysis in Kanekar et al.\cite{2015JPlPh..81a3004K}, we define a kinetic analogue of the Langevin equation by considering the system
\begin{subequations}\label{langsyst}
  \begin{align}
    &\frac{\partial g}{\partial t}+v\frac{\partial g}{\partial x}+vF_0\frac{\partial \varphi}{\partial x}=\chi F_0 \label{eq1lang}\\
     &\varphi=\alpha \int dv\, g. \label{eq2lang}
\end{align}  
\end{subequations}
Both equations are presented in dimensionless form. Specifically, the spatial coordinate $x$ is normalized by an arbitrary length $L$ such that it ranges within the interval $[0,L_{\text{box}}]$, the velocity $v$ is normalized by the thermal speed $v_{\text{th}}=\sqrt{k_B T/m}$ (where $k_B$ is the Boltzmann constant, $T$ is the plasma temperature, and $m$ is the mass of the species under consideration) and time is normalized by $L/v_{\text{th}}$.

Eq.~\eqref{eq1lang} is a linearized Vlasov equation --where we introduced a noise source term whose properties are described below-- describing the evolution of the plasma species in the kinetic regime. In particular, $g(x,v,t)$ is the perturbation to the distribution function $f(x,v,t)$ with respect to the Maxwellian equilibrium $F_0(v)$: $f(x,v,t)=F_0(v)+g(x,v,t)$ with $g/F_0\ll 1$ and $F_0(v)=n_0 e^{-v^2/2}/\sqrt{2\pi}$ ($n_0=1$ in numeric units). The stochastic forcing reads as $\chi(t,x)=a_0 q(x) s(t)$ in such a way that the power injected into the system scales as $\propto a_0^2$. The spatial part of the forcing is defined as:
\begin{equation}
    q(x)= \sum_{n=1}^{N_m}\sqrt{\frac{2}{N_m}}\sin\leri{nk_0x+\phi_n},
\end{equation}
where $N_m$ is the number of modes, $k_0=2\pi/L_{\text{box}}$ is the fundamental harmonic, $\phi_n$ are random phases. It holds that $\int dx\, q(x)=0$ and $\int dx\, q^2(x)=1$. The temporal part $s(t)$ satisfies
\begin{equation}\label{autocorr}
 \langle s(t) s(t')\rangle= e^{-\lambda |t-t'|}
\end{equation}
where  $\langle \cdot \rangle$ represents an ensemble average. The function $s(t)$ encodes the correlated noise driving the turbulence in the kinetic system \cite{Carbone_2022}. The parameter $\lambda=1/\tau_c$, being $\tau_c$ the correlation time, dictates the nature of the forcing term. Specifically, its ratio to the characteristic timescales of the problem (such as the wave frequency or the Landau damping rate) establishes whether the system responds to a highly-correlated forcing or operates in a low-correlation regime. The phenomenology associated with a white-noise source as in Ref. \cite{2015JPlPh..81a3004K} is consistently recovered in the limit $\lambda \to \infty$.  

Eq.~\eqref{eq2lang} links the perturbation in the plasma distribution function with the electrostatic potential $\varphi$. As discussed in Ref. \cite{2015JPlPh..81a3004K}, different physical scenarios can be reproduced by tuning the parameter $\alpha$. For instance, with $\alpha>0$, the system \eqref{langsyst} can represent a Vlasov-Poisson plasma composed by kinetic electrons and motionless ions, describing electronic Langmuir waves. In this case, when Eq.~\eqref{eq2lang} is transformed in Fourier space, the parameter $\alpha$ is expressed as $\alpha=1/(k^2\lambda_D^2)$, being $\lambda_D$ the Debye length and $k$ the wavenumber. We remark that the normalized electric potential $\varphi$ must be defined in terms of the physical one $\phi$ as $\varphi=q\phi/k_BT$, where $q$ is the signed charge of the kinetic species. Therefore, when electrons are considered, the signs of the terms in Eq.~\eqref{eq1lang} and Eq.~\eqref{eq2lang} are consistent with a positive sign for $\alpha$. The case $\alpha>0$ encompasses also ion-acoustic waves. Here $g$ represents the perturbation to the ion distribution functions, while electrons have an instantaneous Boltzmann response, i.e. $n_e=n_0 e^{\varphi}$ where $n_e$ and $n_0$ are the electron and background density, respectively. In this case, $\alpha=ZT_e/T_i$ with $Z$ the ions atomic number, and Eq.~\eqref{eq2lang} can be either derived from the statement of quasi-neutrality or from the Poisson equation in the limit $ZT_e/T_i\gg k^2\lambda_D^2$. Conversely, setting $\alpha<0$ yields a system that represents slow-MHD compressive modes in the high-$\beta$ (see, Ref. \cite{2015JPlPh..81a3004K} for further details). The allowed interval of variation of the parameter $\alpha$ is $(-1,+\infty)$. The singular case $\alpha=-1$ represents an undamped and unforced system, for which the analogy with Langevin phenomenology is no longer feasible. On the other hand, when the $\alpha<-1$ case is considered, the solutions of the dispersion relation correspond to instabilities (not shown here). In this regime, the exponential growth of the perturbations prevents the system from reaching the steady-state equilibrium required to establish an FDR; therefore, this scenario lies outside the scope of the present work.

In the following, we summarize the derivation of the FDR for the system (\ref{langsyst}), by employing the approach presented in Ref.~\cite{2015JPlPh..81a3004K} and extending it to the case of a correlated noise forcing. Applying a Fourier transform\footnote{Here and in the following, we use $k$, by a slight abuse of notation, to denote the dimensionless wavenumber associated with the coordinate $x$ in \eqref{eq1lang}.} in time and space to Eq.~\eqref{eq1lang} provides:
\begin{equation}
    g_{k,\omega}=-\frac{vF_0}{v-\frac{\omega}{k}}\varphi_{k,\omega}-\frac{i F_0}{k\leri{v-\frac{\omega}{k}}}\chi_{k,\omega},
\end{equation}
which, inserted back into Eq.~\eqref{eq2lang} gives a closed expression for the electric potential, namely
\begin{equation}\label{phiFourier}
    \varphi_{k,\omega}=-\frac{i\chi_{k,\omega}}{k}\frac{Z(\zeta)}{D_\alpha(\zeta)}.
\end{equation}
Here, we have introduced the plasma dispersion function and the dielectric function \cite{1973ppp..book.....K}:
\begin{align}
    &Z(\zeta)=\int dv\, \frac{F_0}{v-\zeta}\\
    &D_\alpha(\zeta)=\frac{1}{\alpha}+\int dv\, \frac{vF_0}{v-\zeta}=1+\frac{1}{\alpha}+\zeta Z(\zeta)
\end{align}
where $\zeta=\omega/k$ is the wave phase velocity. The dispersion relation of the collective plasma modes is provided by the pairs $\leri{\omega,k}\in \mathbb{C}\times\mathbb{R}$ satisfying $D_{\alpha}\leri{\omega/k}=0$. In order to evaluate the steady-state spectrum of the electric perturbations, we back-transform Eq.~\eqref{phiFourier} to the time domain, compute its square modulus, and take the ensemble average:
\begin{multline}
\label{spettroprovv}
\langle|\varphi_k|^2\rangle= \frac{1}{\leri{2\pi k }^2}\int d\omega  \int d\omega' \,   \langle\chi_{k,\omega}\chi^*_{k,\omega'} \rangle \times \\
\frac{Z\leri{\frac{\omega}{k}}}{D_\alpha\leri{\frac{\omega}{k}}}\frac{Z^*\leri{\frac{\omega'}{k}}}{D^*_\alpha\leri{\frac{\omega'}{k}}} e^{-i\leri{\omega-\omega'} t }.    
\end{multline}

The noise auto-correlation $\langle\chi_{k,\omega}\chi^*_{k,\omega'} \rangle$ can be evaluated via Fourier analysis and exploiting the forcing temporal property in Eq.~\eqref{autocorr}, resulting in
\begin{equation}\label{chiautocorr}
    \langle \chi_{k,\omega}\chi^*_{k,\omega'} \rangle = 2\pi a_0^2 |q_k|^2\delta(\omega-\omega')\frac{2\lambda}{\lambda^2+\omega^2}.
\end{equation}
Inserting this expression in Eq.~\eqref{spettroprovv} provides the FDR for the correlated noise source case:
\begin{equation}\label{FDRsinglelambda}
    \langle |\varphi_k|^2 \rangle = \frac{ a_0^2 |q_k|^2}{\pi k }\int_{-\infty}^\infty d\zeta \, \frac{\lambda}{\lambda^2+k^2\zeta^2}\left | \frac{Z(\zeta)}{D_\alpha(\zeta)}\right |^2.
\end{equation}

We remark that, by taking the limit $\lambda\to \infty$ of Eq. \ref{FDRsinglelambda}, we recover the FDR relation derived in Ref.~\cite{2015JPlPh..81a3004K} for the white-noise case, that reads as:
\begin{equation}
    \langle |\varphi_k|^2 \rangle = \frac{ \epsilon_k}{2\pi k }\int_{-\infty}^\infty d\zeta \, \left | \frac{Z(\zeta)}{D_\alpha(\zeta)}\right |^2.
\end{equation}
Ref. \cite{2015JPlPh..81a3004K} defines the noise power spectrum $\langle \chi(t)\chi(t')\rangle=\epsilon_k \delta(t-t')$, whereas in our case we assume $\langle \chi(t)\chi(t')\rangle=a_0^2 |q_k|^2 e^{-\lambda|t-t'|}$. In order to find the equivalence between these two different expressions, we calculate, for the two cases, the total energy injected into the system for long times. Equating them, we obtain:
\begin{equation}
    \int_{-\infty}^\infty dt\, \epsilon_k \delta(t) =  \int_{-\infty}^\infty dt\, a_0^2 |q_k|^2 e^{-\lambda|t|} \, ,
\end{equation}
which implies $\epsilon_k=2a_0^2|q_k|^2 \lambda^{-1}$. 
The same result can be achieved by noting that $\delta(t)=\lim\limits_{\lambda \to \infty} \frac{\lambda}{2}e^{-\lambda|t|}$, where the limit is understood in the distributional sense. 

\subsection{Asymptotic scaling}\label{sec2.1}

It is worth calculating asymptotic scalings of Eq.~\eqref{FDRsinglelambda} as these provide interesting insights of the underlying physics in some limit cases. In particular, we will analyze the asymptotic behavior of the spectrum in the limits $\alpha \to -1$ (vanishing phase velocity and weakly damped compressive waves) and $\alpha \to \infty$ (short wavelength weakly damped fast Langmuir waves). As we will show, our findings in the $\alpha \to -1$  and $\alpha \to \infty$ cases reproduce the white-noise phenomenology in the limit of vanishing correlation time, namely $\lambda\to\infty$. Furthermore, we will also consider the short-wavelength limit $k\to\infty$, regardless of the choice of $\alpha$. This limit is equivalent to the highly-correlated regime $\lambda\to 0$ since the relevant ratio characterizing this scenario is $\lambda/k$. 

\subsubsection{Weakly-damped non-oscillatory limit ($\alpha\to -1$)}
When the $\alpha\to -1$ limit is considered, the following approximations hold
\begin{equation}
    Z(\zeta) \approx i\sqrt{\frac{\pi}{2}} \qquad D_\alpha(\zeta)\approx i\sqrt{\frac{\pi}{2}}\leri{\zeta+i \frac{\gamma_L}{k}},
\end{equation}
where the Landau damping rate is given by $\gamma_L=\sqrt{2/\pi}\leri{1+\alpha}k$.
The dispersion relation, calculated from $D_\alpha\leri{\omega/k}=0$, yields a purely imaginary solution, $\omega=-i\gamma_L$, describing a weakly-damped, non-oscillatory solution. In this regime, the FDR takes the approximate form
\begin{equation}
    \langle |\varphi_k|^2 \rangle \approx \frac{ a_0^2 |q_k|^2 \lambda}{\pi k^3 }\int_{-\infty}^\infty d\zeta\, \frac{1}{\zeta^2+\frac{\lambda^2}{k^2}}\frac{1}{\zeta^2+\frac{\gamma_L^2}{k^2}},
\end{equation}
which, through the residue theorem, reads as:
\begin{equation}\label{asymspecmenouno}
     \langle |\varphi_k|^2 \rangle \approx \frac{a_0^2 |q_k|^2}{\gamma_L \leri{\gamma_L+\lambda}} \, .
\end{equation}

Depending on the relative magnitude of the two characteristic rates governing this scenario, namely the damping rate $\gamma_L$ and the correlation frequency $\lambda$, we can identify two distinct behaviors in the asymptotic spectrum. First, when $\lambda \ll \gamma_L$, we observe that $\langle |\varphi_k|^2 \rangle \propto k^{-2}$. In this regime, the damping time is much shorter than the correlation time; this reproduces a highly correlated system where the Landau damping effectively acts on a noise that varies slowly compared to the decay timescale. In the opposite limit, $\lambda \gg \gamma_L$, the noise is practically uncorrelated on the damping timescale, yielding a spectral scaling $\langle |\varphi_k|^2 \rangle \propto k^{-1}$, reproducing the white-noise case \cite{2015JPlPh..81a3004K}. 

\subsubsection{Weakly-damped fast-wave limit ($\alpha \to \infty$)}
In this limit the following approximation holds:
\begin{equation}
    \begin{split}
        &Z(\zeta) \approx -\frac{1}{\zeta}-\frac{1}{\zeta^3}+i\sqrt{\frac{\pi}{2}}e^{-\zeta^2/2} \\ &D_\alpha(x)\approx \frac{1}{\alpha}-\frac{1}{\zeta^2}+i\zeta\sqrt{\frac{\pi}{2}}e^{-\zeta^2/2}.
    \end{split}
\end{equation}
The collective plasma modes obey the dispersion relation
\begin{equation}\label{reldisp2}
    \omega=\pm \sqrt{\alpha}k-i \gamma_L \qquad \gamma_L=\sqrt{\frac{\pi}{8}}\alpha^2 e^{-\alpha/2}k,
\end{equation}
describing Langmuir waves with phase velocity much greater than the thermal speed. In this regime, Landau damping is exponentially small, allowing these modes to persist over long timescales within the plasma. Under these conditions the plasma response is well approximated by the sum of two narrow Lorentzians:
\begin{equation}
    \left|\frac{Z(\zeta)}{D_\alpha(\zeta)}\right|^2 \approx \frac{\alpha^2}{4}\leri{\frac{1}{\leri{\zeta-\sqrt{\alpha}}^2+\frac{\gamma_L^2}{k^2}}+\frac{1}{\leri{\zeta+\sqrt{\alpha}}^2+\frac{\gamma_L^2}{k^2}}}.
\end{equation}
By inserting last expression into the integral appearing in Eq.~\eqref{FDRsinglelambda} and applying the residue theorem, we obtain two separate contributions, one arising from the plasma poles $\mathcal{I}_P$ and the other related to the noise $\mathcal{I}_N$, reading
\begin{align}
    &\mathcal{I}_{P}=\frac{\lambda\pi \alpha^2 k}{2\gamma_L\leri{\alpha k^2+\lambda^2}} \\
    &\mathcal{I}_N=\frac{\pi\alpha^2 k \leri{\alpha k^2-\lambda^2}}{2\leri{\alpha k^2+\lambda^2}^2} \, ,
\end{align}
whose ratio is: 
\begin{equation}
    \frac{\mathcal{I}_P}{\mathcal{I}_N}=\frac{\lambda\leri{\alpha k^2+\lambda^2}}{\gamma_L\leri{\alpha k^2-\lambda^2}}.
\end{equation}
The plasma term is hence exponentially large with respect to the noise one (see Eq.~\eqref{reldisp2}), therefore the latter can be safely neglected, yielding this simple expression for the asymptotic spectrum in the large-$\alpha$ limit:
\begin{equation}\label{asymspecinfinity}
    \langle |\varphi_k|^2 \rangle \approx \frac{\lambda a_0^2 |q_k|^2\alpha^2}{2\gamma_L \leri{\alpha k^2+\lambda^2}}.
\end{equation}
Two separate scenarios emerge depending on the relative magnitude of the correlation frequency $\lambda$ and the wave frequency $\omega_L=\sqrt{\alpha}k$. In the highly-correlated case, i.e. $\lambda \ll \omega_L$ or, in other words, correlation time much larger than the Langmuir wave period, we expect a spectral scaling $\langle |\varphi_k|^2 \rangle \propto k^{-3}$. Conversely, in the white-noise limit $\lambda \gg \omega_L$, we recover the standard scaling outlined in \cite{2015JPlPh..81a3004K}, namely $\langle |\varphi_k|^2 \rangle \propto k^{-1}$.  

\subsubsection{Short-wavelength limit ($k\to\infty$)}
As previously stated, the short-wavelength limit, performed without further assumptions on $\alpha$, is of great importance in our analysis. The electrostatic kinetic framework developed here is inherently best suited to describe plasma turbulence at the smallest spatial scales, corresponding to the sub-ion and sub-electron regimes (i.e., beyond the spectral break) of solar wind turbulence. 
At such small scales the fluctuation spectrum becomes universally governed by the static (zero-frequency) susceptibility of the plasma. Given that the relevant dynamical parameter is the ratio $\lambda/k$, the limit $k \to \infty$ intrinsically mimics the $\lambda \to 0$ limit. Consequently, at the smallest spatial scales, the plasma always perceives the forcing as a highly-correlated noise, regardless of the actual finite value of the correlation rate $\lambda$.

The mathematical treatment of this limit is straightforward. By rewriting the FDR as
\begin{equation}
 \langle |\varphi_k|^2 \rangle  = \frac{a_0^2|q_k|^2}{\pi k^3 }\int_{-\infty}^\infty d\zeta \, \frac{\lambda}{\frac{\lambda^2}{k^2}+\zeta^2}\left | \frac{Z(\zeta)}{D_\alpha(\zeta)}\right |^2,
\end{equation}
we can perform a distributional limit from the Lorentzian to a Dirac delta. The approximate form for the FDR is readily obtained
\begin{equation}
    \langle |\varphi_k|^2 \rangle \approx\frac{ a_0^2 |q_k|^2}{k^2 }\int_{-\infty}^\infty d\zeta \, \delta(\zeta)\left | \frac{Z(\zeta)}{D_\alpha(\zeta)}\right |^2=\frac{ \pi a_0^2 |q_k|^2\alpha^2}{2 k^2 \leri{\alpha+1}^2}.
\end{equation}

This mathematical outcome carries a precise physical meaning. At asymptotically small scales, the phase velocity of the perturbations vanishes ($\omega/k \to 0$), and the plasma loses its ability to sustain collective wave modes. Consequently, dynamical kinetic effects are entirely suppressed, and the system responds solely through static screening, encoded in the $\alpha$-dependent pre-factor. This implies that the transition to the $k \to \infty$ regime effectively acts as a kinetic spectral break. At scales smaller than such a break, wave-particle interactions cease to actively shape the fluctuation spectrum. Since the electrostatic potential scales as $\langle|\varphi_k|^2\rangle \propto k^{-2} |q_k|^2$, the corresponding electric field fluctuations, obtained via $\langle|E_k|^2\rangle = k^2 \langle|\varphi_k|^2\rangle$, passively mirror the spatial injection spectrum, yielding $\langle|E_k|^2\rangle \propto |q_k|^2$. At these extreme scales, the electric field response becomes entirely uncoupled from the temporal correlation parameter $\lambda$.

\subsection{Hermite decomposition}
To elucidate the features of the turbulent cascade in velocity space, it is highly instructive to perform a Hermite decomposition of the perturbed distribution function $g(x,v,t)$. We make use of the probabilist's Hermite polynomials, defined as $\text{He}_n(v)=(-1)^n e^{v^2/2}\frac{d^n}{dv^n}e^{-v^2/2}$. The projection onto the Hermite basis and its corresponding inversion formula are given by
\begin{equation}
\begin{split}
 &g_m = \int_{-\infty}^\infty dv\, \frac{\text{He}_m(v)}{\sqrt{ m!}}g(v) \\
 &g(v) = \sum_{m=0}^\infty \frac{\text{He}_m(v) F_0(v)}{\sqrt{ m!}}g_m.
    \end{split}
\end{equation}
By applying the Hermite transform to the Eq.~\eqref{eq1lang} and following the same steps that led to Eq.~\eqref{FDRsinglelambda}, we can derive the steady-state Hermite spectrum:
\begin{multline}
    C_{m,k}\equiv \left \langle \left | g_{m,k}\right |^2\right \rangle= \frac{ a_0^2 |q_k|^2}{\pi k m!}\leri{\frac{\alpha+1}{\alpha}}^2 \times \\
    \int_{-\infty}^\infty d\zeta\, \frac{\lambda}{\lambda^2+k^2\zeta^2} \left |\frac{Z^{(m)}(\zeta)}{D_{\alpha}(\zeta)} \right |^2,
\end{multline}
where we have defined
\begin{equation}
    Z^{(m)}\equiv \frac{d^mZ}{d\zeta^m}=\leri{-1}^m\int_{-\infty}^\infty dv\, \frac{\text{He}_m(v) F_0(v)}{v-\zeta}.
\end{equation}
The asymptotic behavior of this spectrum for large $m$ is known to scale as $C_{m,k}\approx A_k(\alpha,\lambda)/\sqrt{m}$ (see Ref.\cite{2015JPlPh..81a3004K} and references therein), where $A_k$ is:
\begin{multline}\label{ampiezza}
A_k(\alpha,\lambda)=\frac{\lambda a_0^2 |q_k|^2}{\sqrt{2\pi}k}\leri{\frac{\alpha+1}{\alpha}}^2 \times \\
\int_{-\infty}^\infty d\zeta\, \frac{e^{-\zeta^2/2}}{\leri{\lambda^2+k^2\zeta^2}|D_{\alpha}(\zeta)|^2} .
\end{multline}
By implementing the same strategy proposed in Ref. \cite{2015JPlPh..81a3004K}, we evaluate the integral contained in the above formula: 
\begin{equation}\label{ampiezzahermite}
A_k\leri{\alpha,\lambda}=\frac{a_0^2 |q_k|^2}{\lambda k}\leri{\frac{\alpha+1}{\alpha}}^2\leri{\frac{1}{D_\alpha\leri{i\lambda/k}}-\frac{\alpha}{1+\alpha}},
\end{equation}
where the plasma dispersion function evaluated on the imaginary axis is:
\begin{equation}
    D_\alpha\leri{i\lambda/k}=1+\frac{1}{\alpha}-\sqrt{\frac{\pi}{2}}\frac{\lambda}{k} e^{\frac{\lambda^2}{2k^2}}\text{erfc}\leri{\frac{\lambda}{\sqrt{2}k}}.
\end{equation}

In order to demonstrate that we consistently recover the white-noise phenomenology in this limit, it is sufficient to recall that the complementary error function $\text{erfc}(y)$ admits the following asymptotic expansion for large values of its argument:
\begin{equation}
    \text{erfc}(y)\approx\frac{e^{-y^2}}{y\sqrt{\pi}}\left(1-\frac{1}{2y^2}\right).
\end{equation}
By setting $y=\lambda/(\sqrt{2}k)$ and taking the limit $\lambda \to \infty$, we obtain the asymptotic amplitude:
\begin{equation}\label{amplimkan}
    A_k\left(\alpha,\lambda\right)\approx \frac{\sqrt{2} a_0^2|q_k|^2}{\lambda k}\left(1+\alpha\right).
\end{equation}
By recalling the formal equivalence for the injected power $\epsilon_k = 2 a_0^2 |q_k|^2 / \lambda$ derived earlier, we immediately recognize that Eq.~\eqref{amplimkan} yields $A_k \approx \epsilon_k(1+\alpha)/(\sqrt{2}k)$. This is precisely the cascade amplitude associated with a white-noise driven system \cite{2015JPlPh..81a3004K}. This confirms that our generalized model accurately recovers the classic white-noise phenomenology when the turbulent forcing decorrelates instantaneously.

It is important to remark that both the $k$-space and Hermite steady-state spectra describe local, rather than global, equilibria, i.e. they accurately characterize the system at fixed $k$ and $m$ over a suitable timescale. In other words, energy is continuously injected in the system by the source term at $m=0$ (the noise is Maxwellian in the velocity space) and transferred to higher Hermite modes via Landau damping, therefore equilibrium is reached at a given $m$ only after the energy cascade in Hermite space has had sufficient time to reach that particular mode.

\section{Numerical simulations}
\label{sec3}

In this section, we test the analytical predictions derived above against numerical simulations. We perform direct integrations of the equations system, Eqs.~\eqref{langsyst}, by employing a spectral algorithm. We assume periodic boundary conditions on the spatial coordinate $x$ and evaluate spatial derivatives by means of a spectral approach. We set $L_{\text{box}}=4\pi$ and use a grid of $N_x=256$ evenly spaced points covering the spatial domain. The grid in the velocity space consists of $N_v=2001$ points on the segment $[-6,6]$ (we recall that velocities are normalized to the thermal speed). We remark that no boundary conditions must be imposed on the velocity $v$, for the fact that the latter enters the linearized system Eqs.~\eqref{langsyst} as a parameter, rather than a dynamical variable. In all simulations, we assume a forcing amplitude $a_0=10^{-3}$ and a superposition of $N_m = 8$ independent modes for the spatial part of the noise, while the phases $\phi_n$ are randomly sampled for each run. Time integration is performed using a standard fourth-order Runge-Kutta scheme. The white-noise case results have been preliminarily reproduced in order to validate the performances of our numerical algorithm before applying it to the finite correlation time case. 

We avoid introducing any numerical dissipation term as, for example, small collisional term or hyperviscosity for preventing eventual numerical instabilities. On one hand, this ensures that Landau damping is indeed the only energy conversion process at play in balancing the power supplied to the system by the random source, thus perfectly mimicking the analytical FDR derived above. On the other hand, this implies that other strategies are needed to prevent numerical artifacts. In this perspective, we set our maximum simulation time $T_{\text{max}}$ and the velocity-space grid size $\Delta v$ in such a way that $T_{\text{max}} < T_{\text{rec}}$ is satisfied for all driven modes $\{k_1, \dots, k_{N_m}\}$. Here, $T_{\text{rec}} = 2\pi/(k\Delta v)$ is the free-streaming recurrence time \cite{2016PhPl...23b2103P}, which marks the onset of artificial echoes due to the discreteness of the velocity grid. The time step $\Delta t$ is also properly adjusted for each case depending on the value of $\lambda$ to ensure that the numerical solution appreciates the forcing decorrelation $\sim 1/\lambda$: $\Delta t\ll1/\lambda$.

\begin{figure}
\centering
\includegraphics[width=8cm]{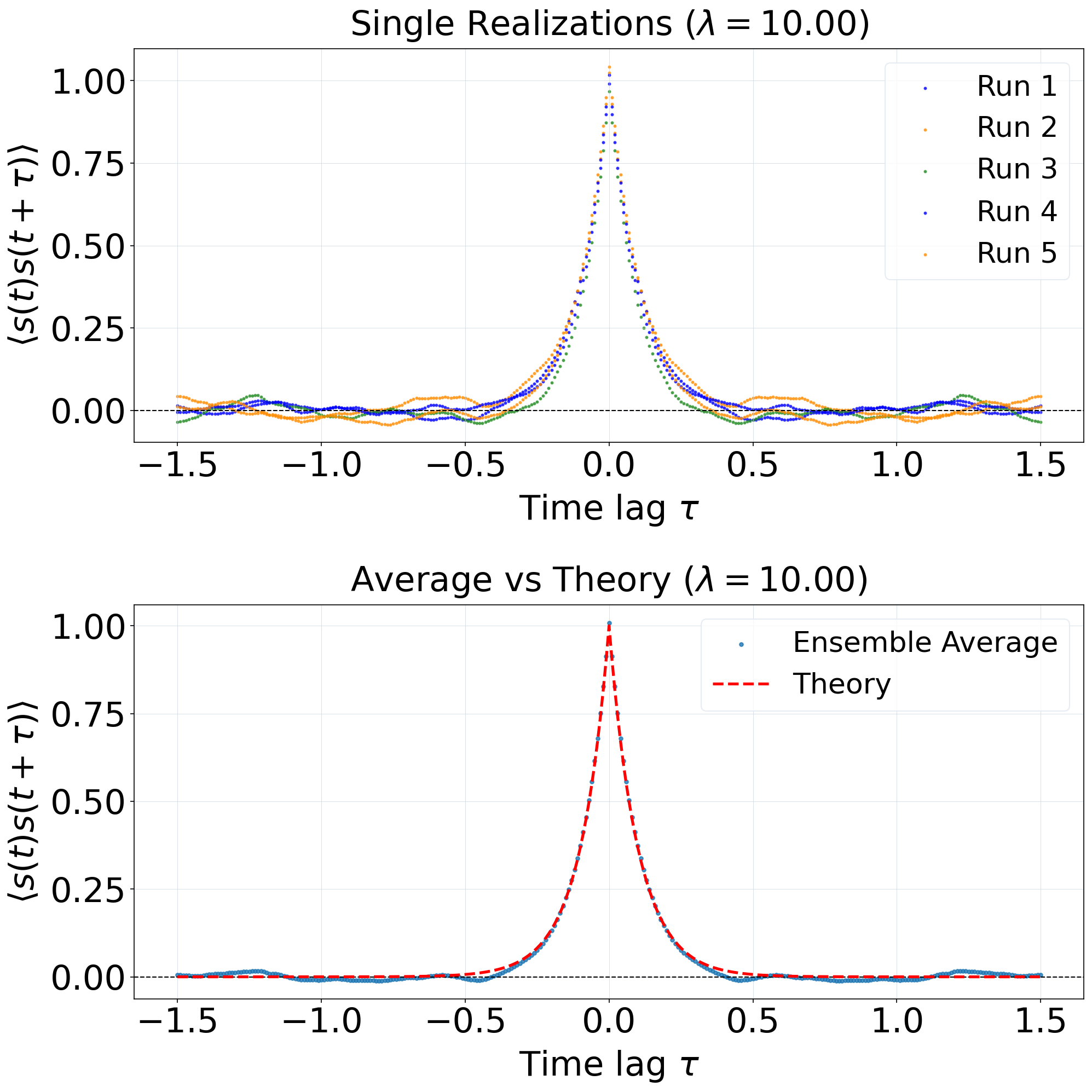}
\caption{Convergence of the numerically generated correlated noise to the theoretical curve. \textit{Upper panel}: independent realizations. \textit{Lower panel}: ensemble average and theoretical curve. }
\label{fig:autocorr}
\end{figure}

A final comment on the forcing term is in order. The correlated noise is generated through an Ornstein–Uhlenbeck process \cite{PhysRev.36.823}, which is inherently stochastic. For each case, the specific numerical realization of the noise is slightly affected by random fluctuations, and the convergence toward the expected analytical shape of the autocorrelation function is achieved only by ensemble averaging over an appropriate number of runs. As appreciated in Fig.~\ref{fig:autocorr}, five independent realizations of the numerical noise for a given set of the relevant parameters ($\lambda$, $T_{\text{max}}$, and $dt$) show slight discrepancies with respect to the theoretical behavior (red dashed line), while their average smoothly follows it (light blue points). Given this premise, it is clear that the numerical solution of Eqs.~\eqref{langsyst} will be affected by the same random fluctuations that characterize the source term. Consequently, all numerical verifications of the analytical FDRs reported in this section can only be achieved by averaging over a sufficient number of independent realizations. For each realization, the forcing drives the growth of the electrostatic potential. In particular, the energy of the driven modes reaches a saturation regime at which we employ the FDR analysis, as detailed below. Note also that, for each realization, the invariant conservations is excellent: for example, the Boltzmann-Gibbs entropy is preserved up to $10^{-3} \%$. 

As a first testcase, we present the numerical verification of the FDR provided in Eq.~\eqref{FDRsinglelambda}. The procedure consists of: $i)$
simulating Eqs.~\eqref{langsyst}, $ii)$ calculating, from the output of the numerical integration, the steady-state spectrum $\langle |\varphi_k|^2\rangle_{\text{num}}$ where the average is computed on a time window of size $T_{\rm aver}=250$ in the saturation regime, and $iii)$ ensemble-averaging the results over a certain number of trials $N_{\text{ens}}$. The numerical verification of the FDR is then achieved if
\begin{equation}\label{fdrverify}
    \frac{\langle |\varphi_k|^2 \rangle_{\text{num}}}{\frac{ a_0^2 |q_k|^2}{\pi k }\int_{-\infty}^\infty d\zeta \, \frac{\lambda}{\lambda^2+k^2\zeta^2}\left | \frac{Z(\zeta)}{D_\alpha(\zeta)}\right |^2} = 1
\end{equation}
holds for all driven $k$. We remark again that, in each run, specific values for $\alpha$ and $\lambda$ must be chosen and the denominator in the above equation is calculated separately from the numerical integration of Eqs.~$\eqref{langsyst}$. 

\begin{figure}[hbtp!]
\centering
\includegraphics[width=8cm]{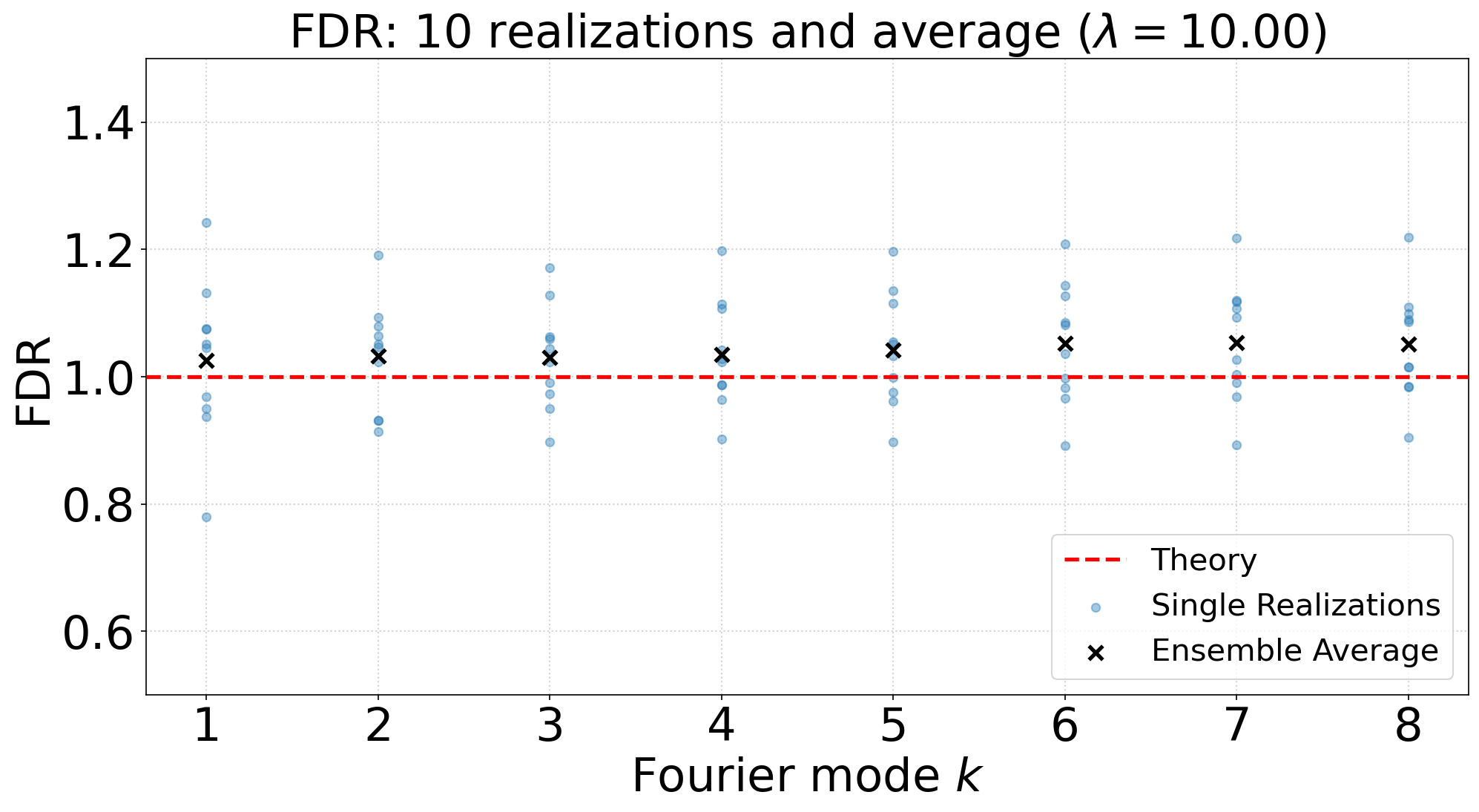}
\caption{Numerical verification of the FDR. The light blue dots indicate the 10 independent realizations, the black crosses represent the ensemble average.}
\label{fig:fdrverify}
\end{figure}

In Fig.~\ref{fig:fdrverify}, we present the outputs of $N_{ens}=10$ integrations of Eqs.~\eqref{langsyst} in which we set $\lambda=10$, $\alpha=1$, and $\Delta t=0.05$. The plot clearly shows that the fluctuations, arising from the stochastic nature of the forcing, are symmetrically distributed around the expected theoretical behavior. On the other hand, by looking at the black crosses describing the mean obtained from the independent realizations, the mean deviation from the expected unity value is below $5\%$ for the chosen ensemble size.

\begin{figure}[hbtp!]
\centering
\includegraphics[width=8cm]{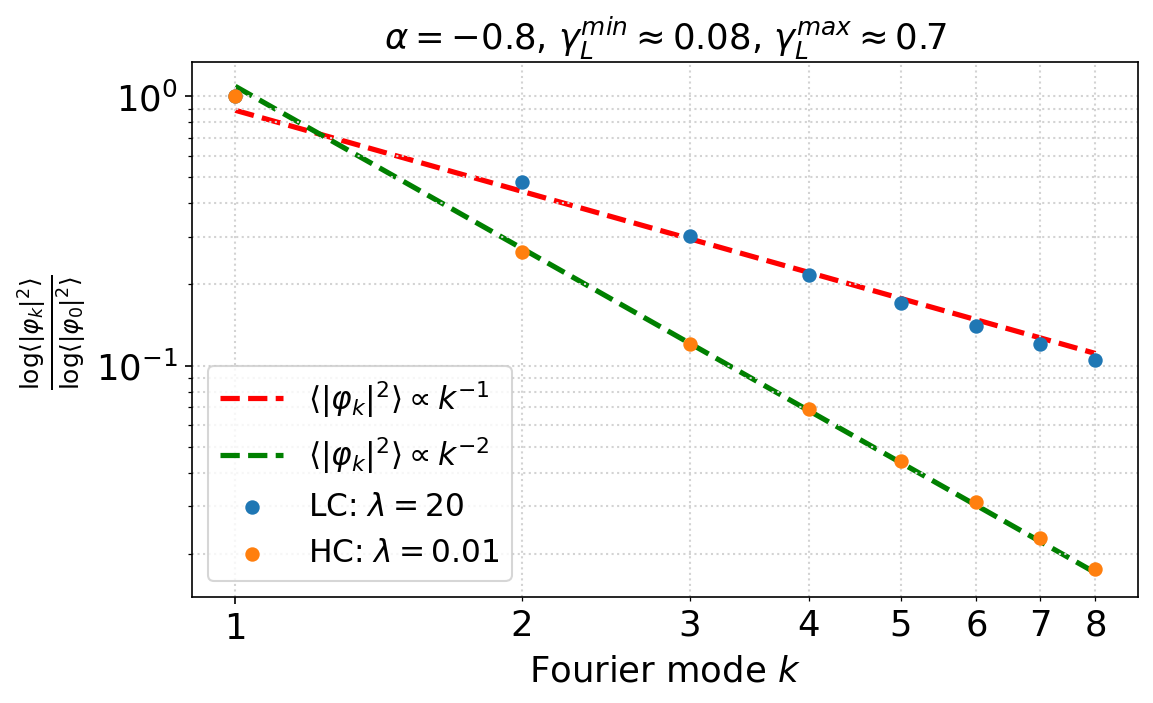}
\caption{Numerical verification of the asymptotic scaling in the $\alpha \to -1$ limit, simulated with $\alpha = -0.8$. The low correlation (LC) case, i.e. the regime $\lambda\gg\gamma_L^{\text{max}}=0.7$, is reproduced by setting $\lambda=20$. Conversely, the high correlation (HC) case, i.e. the regime $\lambda\ll\gamma_L^{\text{min}}=0.08$, is reproduced by setting $\lambda=0.01$.}
\label{fig:scalingmenouno}
\end{figure}

Then, we present the numerical verification of the asymptotic scalings analytically derived in Sec.~\ref{sec2.1}. We start with the case $\alpha \to -1$, which we reproduce by setting $\alpha=-0.8$. Choosing this value for $\alpha$ provides a range of variation for the Landau damping rate between $\gamma_L^{\text{min}}=0.08$ and $\gamma_L^{\text{max}}=0.7$. The two regimes of high and low correlation are achieved by setting $\lambda=0.01$ and $\lambda=20$, respectively. In this limit, Landau damping vanishes linearly; however, for $\alpha=-0.8$, it still provides an effective mechanism for energy transfer towards the small scales. As displayed in Fig.~\ref{fig:scalingmenouno}, the predicted theoretical behavior is well reproduced even when averaging over a small ensemble of runs $N_{\text{ens}}=5$.

\begin{figure}[hbtp!]
\centering
\includegraphics[width=8cm]{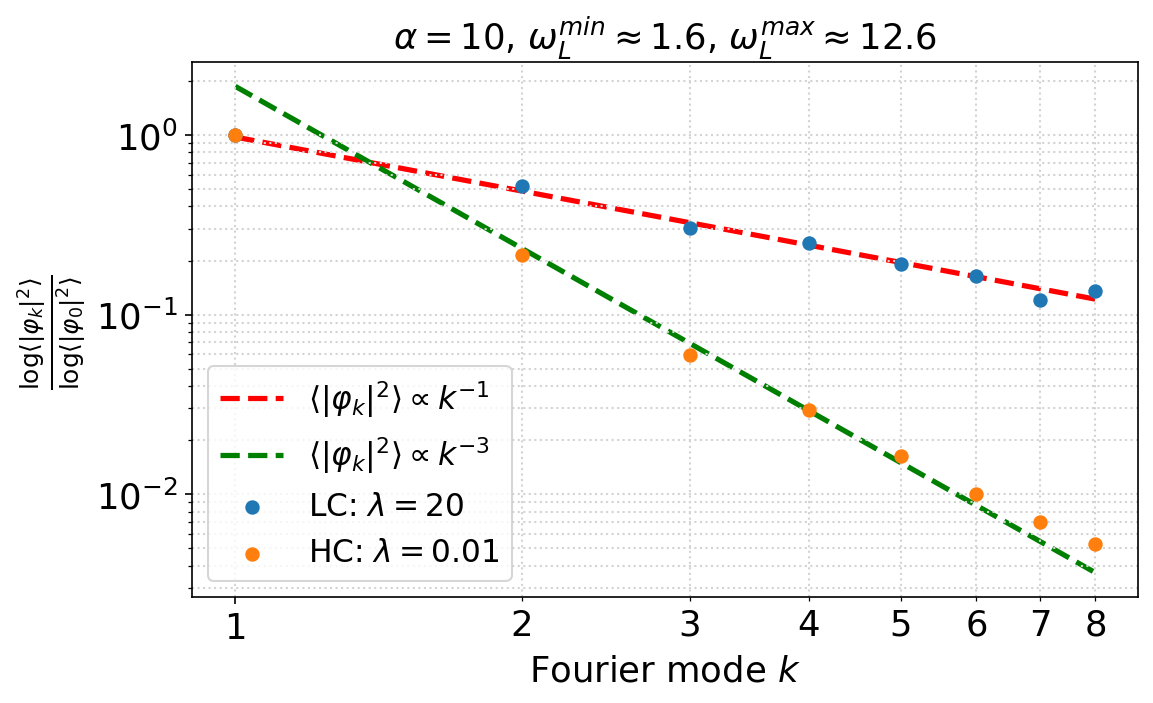}
\caption{Numerical verification of the asymptotic scaling in the $\alpha \to \infty$ limit, simulated with $\alpha = 10$. The low correlation (LC) case , i.e. the regime $\lambda\gg\omega_L^{\text{max}}\approx12.6$, is reproduced by setting $\lambda=150$. Conversely, the high correlation (HC) case, i.e. the regime $\lambda\ll\omega_L^{\text{min}}\approx1.6$, is reproduced by setting $\lambda=0.1$.}
\label{fig:scalinginfinito}
\end{figure}

Instead, in the $\alpha\to \infty$ limit, the Landau damping rate vanishes exponentially. By setting $\alpha=10$, the energy transfer across scales is very weak, and the integration of Eqs.~\eqref{langsyst} is more affected by numerical noise. Nevertheless, with $N_{\text{ens}}=20$ realizations, a very good agreement with the theoretical predictions can still be achieved, as shown in Fig.~\ref{fig:scalinginfinito}. Here, the cases of high and low correlation are given by $\lambda=0.1\ll\omega_L^{\text{min}}\approx 1.6$ and $\lambda=150\gg\omega_L^{\text{max}}\approx 12.6$, respectively.

\begin{figure}[hbtp!]
\centering
\includegraphics[width=8cm]{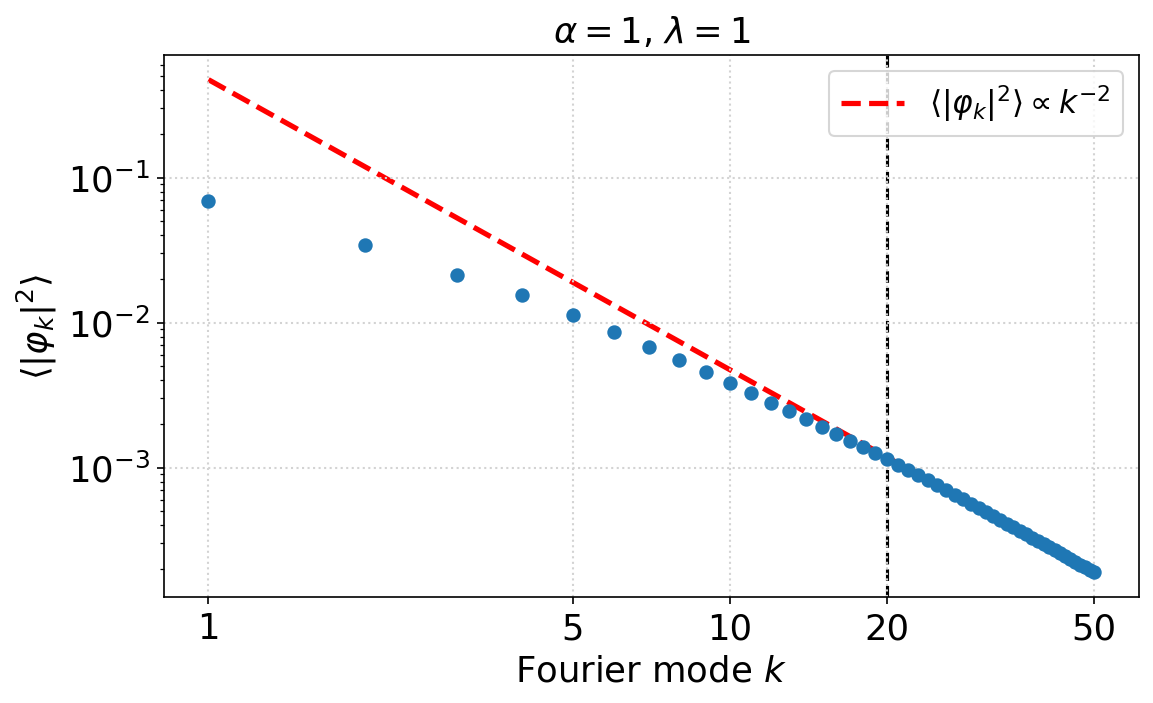}
\caption{Numerical verification of the short-wavelength asymptotic scaling ($k \to \infty$). By expanding the spatial domain to $L_{\text{box}}=40\pi$ and forcing $N_m=50$ modes, the transition from the $\lambda/k \gg 1$ regime to the short-scale limit ($\lambda/k \ll 1$) is captured within a single simulation. The spectrum smoothly transitions to the predicted $k^{-2}$ scaling around the mode $k_{20}$.}
\label{fig:scalingkgrande}
\end{figure}

Finally, to properly investigate the short-wavelength limit ($k\to\infty$), we adopt a specifically designed numerical setup. First, we increase the number of driven modes to $N_m = 50$ and the spatial grid to $N_x = 512$ points to prevent aliasing at high wavenumbers. Resolving the fast dynamics associated with large $k$ values requires a high time resolution, and, to this purpose, we set $\Delta t=0.001$, while fixing $\alpha=1$ and $\lambda=1$. We also expand the physical domain to $L_{\text{box}}=40\pi$ to ensure that only a specific subset of high-$k$ modes falls into the $\lambda/k\ll 1$ regime, allowing us to clearly observe where the asymptotic scaling $\langle|\varphi_k|^2\rangle\propto k^{-2}$ accurately captures their behavior. The results of this numerical validation are illustrated in Fig.~\ref{fig:scalingkgrande}. The plot clearly demonstrates that the transition from the $\lambda/k\gg 1$ regime to the $k^{-2}$ spectral scaling occurs around the mode $k_{20}=20 k_0$.

\begin{figure}[hbtp!]
\centering
\includegraphics[width=8cm]{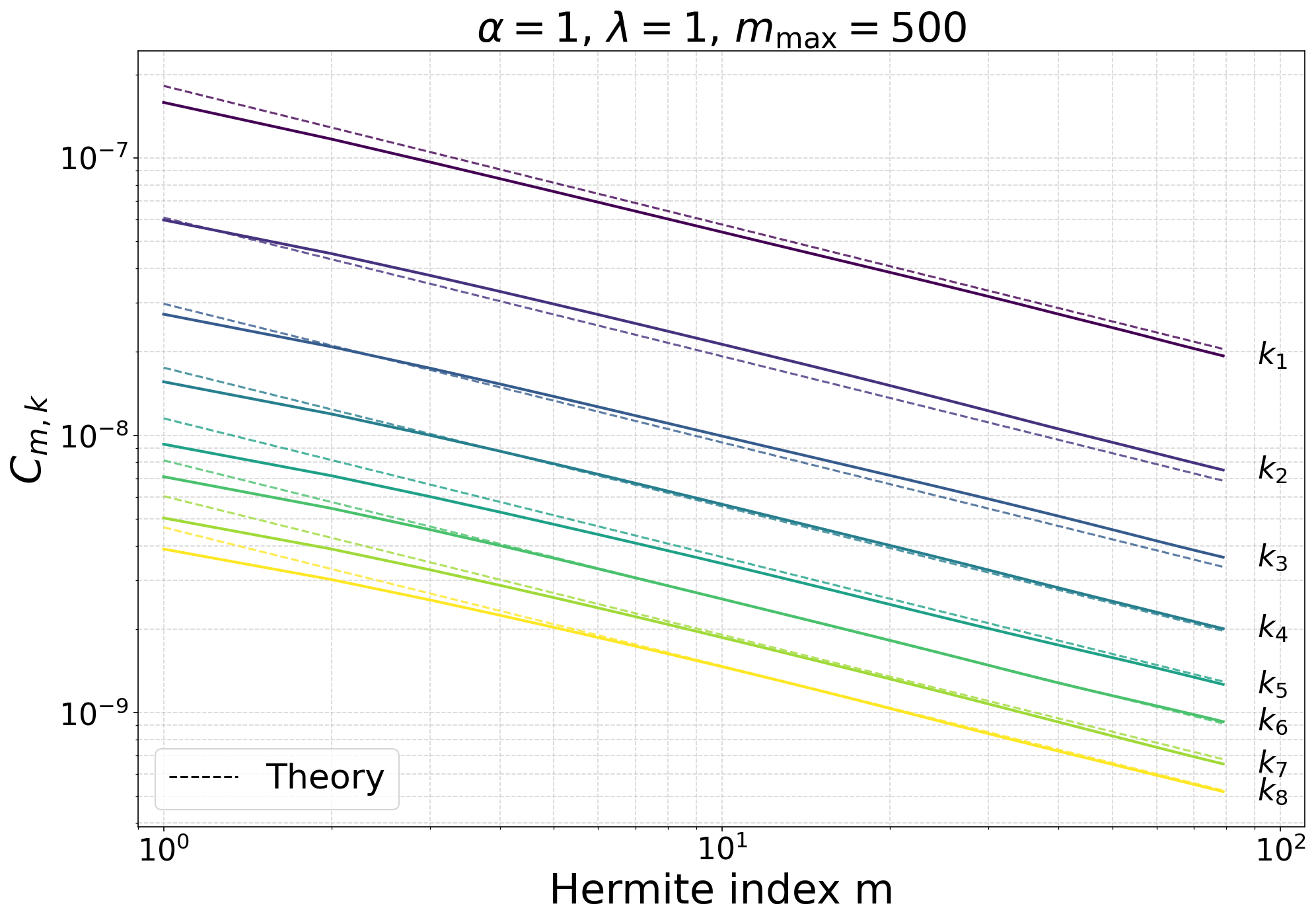}
\caption{Numerical verification of the Hermite steady-state spectrum asymptotic scaling for the all the driven wavenumbers. The numerical data align with the theoretically predicted $A_k/\sqrt{m}$ scaling (dashed lines).}
\label{fig:scalinghermite}
\end{figure}

As a final step in this section, we provide a numerical validation of our analytical findings regarding the steady-state behavior of the Hermite spectrum $C_{m,k}$. For this particular run, the system parameters are $\alpha = 1$ and $\lambda = 1$. The numerical solution of Eqs.~\eqref{langsyst} was projected onto the Hermite basis up to $m_{\text{max}} = 500$ to ensure an adequate resolution of the kinetic dynamics. The corresponding Hermite spectrum for all the wavenumbers included in the forcing term is illustrated in Fig.~\ref{fig:scalinghermite}. The plot reveals a clear and robust agreement between the numerical integration (solid lines) and the theoretically predicted $A_k/\sqrt{m}$ asymptotic scaling (dashed lines) in terms of both the spectral slope and amplitude, the latter being calculated from  Eq.~\eqref{ampiezzahermite}, thus effectively demonstrating the capability of the model to capture fine-scale features in velocity space.

\section{Power-law distribution and link with in-situ observations}
\label{sec4}

A natural extension of the analysis provided in the previous sections is to treat the parameter $\lambda$ as a stochastic variable itself. Indeed, as shown in \cite{Carbone_2022, Chiappetta_2023} by employing Cluster and MMS data, $(i)$ plasma turbulence in near-Earth environments can be appropriately described by a Langevin system where the forcing is a correlated noise, and $(ii)$ the correlation times are distributed according to a power law distribution:
\begin{equation}
dP(\lambda)=C\lambda^{-\mu}d\lambda \, ,
\end{equation}
where $\lambda$ lies in the interval $[\lambda_{\min}, \lambda_{max}]$. The normalization constant $C$ is such that $\int_{\lambda_\text{min}}^{\lambda_\text{max}}C\lambda^{-\mu}d\lambda=1$, $ C=(1-\mu)/(\lambda_\text{max}^{1-\mu}-\lambda_\text{min}^{1-\mu})$.

Assuming a superposition of correlation rates to characterize the turbulence source is a physically well-grounded assumption. In solar wind turbulence, for example, the energy injection from fluid scales to kinetic ones can be understood as the result of multiple processes. Each of these occurs on a characteristic timescale, originating either from the chaotic superposition of convective motions on the solar surface or from the subsequent flow evolution during the propagation. Therefore, the spectral content of solar wind turbulence can be analyzed as the superposition of multiple contributions. By employing model-based fitting procedures, such as those introduced in \cite{2006JCoPh.217..834M}, the most likely distribution of correlation times that reproduces the observed spectra can, in principle, be extracted from the data. Consequently, the observed spectral signatures can be interpreted as the collective response of the plasma to this multiscale stochastic forcing.

Deriving an appropriate version of the FDR for this case is immediate: the only modification with respect to the calculations presented in Sec.~\ref{sec2} occurs when evaluating the noise autocorrelation $\langle \chi_{k,\omega}\chi^*_{k,\omega'} \rangle$ in Eq.~\eqref{chiautocorr}. Now $\lambda$ is a stochastic variable and must be therefore averaged over its probability distribution to obtain a deterministic prediction:
\begin{equation}
    \langle \chi_{k,\omega}\chi^*_{k,\omega'} \rangle = 4\pi C a_0^2 |q_k|^2\delta(\omega-\omega')\int_{\lambda_{\text{min}}}^{\lambda_{\text{max}}} d\lambda\, \frac{\lambda^{1-\mu}}{\lambda^2+\omega^2},
\end{equation}
 which, plugged into the electric potential spectrum, gives the kinetic FDR for a power-law distributed correlation time of the noise source:
\begin{multline}\label{FDRaverage}
     \langle |\varphi_k|^2 \rangle = \frac{C a_0^2 |q_k|^2}{\pi k }\times \\
\int_{\lambda_{\text{min}}}^{\lambda_{\text{max}}} d\lambda \, \lambda^{1-\mu}\int_{-\infty}^\infty d\zeta\, \frac{1}{\lambda^2+k^2\zeta^2} \left | \frac{Z(\zeta)}{D_\alpha(\zeta)}\right |^2.
\end{multline}

The asymptotic spectral scalings derived in Sec.~\ref{sec2.1} in the limits $\alpha\to -1$ and $\alpha \to \infty$ given by Eq.~\eqref{asymspecmenouno} and Eq.~\eqref{asymspecinfinity} must be averaged over the power-law distribution, given their explicit dependence on $\lambda$ (the asymptotic spectrum in the short-wavelength case is independent from $\lambda$, therefore the average and the single-$\lambda$ FDR coincide). When the $\alpha\to -1$ limit is considered, we obtain:
\begin{equation}\label{fdraveragemenouno}
     \langle |\varphi_k|^2 \rangle = \frac{C a_0^2 |q_k|^2}{\gamma_L }\int_{\lambda_\text{min}}^{\lambda_\text{max}} d\lambda \, \frac{\lambda^{-\mu}}{\gamma_L+\lambda}
\end{equation}
and an explicit solution for the integral can be found in terms of the Gaussian hypergeometric function $\prescript{}{2}{F}_1(a, b; c; z)$, giving
\begin{multline}
    \langle |\varphi_k|^2 \rangle = \frac{C a_0^2 |q_k|^2}{\gamma_L^2\leri{1-\mu} } \times\\
    \left.\lerisq{\lambda^{1-\mu}\prescript{}{2}{F}_1 \leri{1, 1-\mu; 2-\mu; -\frac{\lambda}{\gamma_L}}}\right |_{\lambda_\text{min}}^{\lambda_\text{max}} \, .
\end{multline}
The above expression acquires a simplified form by assuming, as suggested in Ref.\cite{Carbone_2022},  $\mu=3/2$. In this specific case we obtain

\begin{multline}\label{closedexpr1}
    \langle |\varphi_k|^2 \rangle = \frac{2C a_0^2 |q_k|^2}{\gamma_L^2}\times \\
\lerisq{\frac{\sqrt{\lambda_\text{max}}-\sqrt{\lambda_\text{min}}}{\sqrt{\lambda_\text{min}\lambda_\text{max}}}+\frac{1}{\sqrt{\gamma_L}}\arctan\leri{\frac{\sqrt{\frac{\lambda_\text{min}}{\gamma_L}}-\sqrt{\frac{\lambda_\text{max}}{\gamma_L}}}{1+\frac{\sqrt{\lambda_\text{max}\lambda_\text{min}}}{\gamma_L}}}}.
\end{multline}

In the limit $\alpha\to\infty$, the asymptotic spectrum averaged over the power-law distribution is:
\begin{equation}\label{fdraverageinfinito}
     \langle |\varphi_k|^2 \rangle = \frac{C a_0^2 |q_k|^2\alpha^2}{2\gamma_L }\int_{\lambda_\text{min}}^{\lambda_\text{max}} d\lambda \, \frac{\lambda^{1-\mu}}{\alpha k^2+\lambda^2} \, .
\end{equation}
Also in this case, the integral in the above formula can be expressed in terms of the Gaussian hypergeometric function:
\begin{multline}
    \langle |\varphi_k|^2 \rangle = \frac{C a_0^2 |q_k|^2\alpha}{2k^2\gamma_L \leri{2-\mu}} \times \\
    \left.\lerisq{\lambda^{2-\mu}\prescript{}{2}{F}_1 \leri{1, 1-\frac{\mu}{2}; 2-\frac{\mu}{2}; -\frac{\lambda^2}{\alpha k^2}}}\right |_{\lambda_\text{min}}^{\lambda_\text{max}}.
\end{multline}
It is interesting to point out that, by setting again $\mu=3/2$, a remarkably simple formula can be calculated:

\begin{multline}\label{closedexpr2}
    \langle |\varphi_k|^2 \rangle = \frac{C a_0^2 |q_k|^2\alpha^2}{2\sqrt{2}\gamma_L \omega_L^{3/2}}\times \\
    \left.\lerisq{\operatorname{arctanh} \leri{\frac{\sqrt{2\lambda\omega_L}}{\lambda+\omega_L}}+\operatorname{arctan} \leri{\frac{\sqrt{2\lambda\omega_L}}{\omega_L-\lambda}}}\right |_{\lambda_\text{min}}^{\lambda_\text{max}},
\end{multline}
where we recall that  $\omega_L=\sqrt{\alpha}k$.

At this stage, it is worth noting the formal and physical analogy between the $\lambda$-averaged spectra obtained here, specifically in Eqs.~\eqref{FDRaverage}, \eqref{fdraveragemenouno} and \eqref{fdraverageinfinito}, and the mathematical framework described in Ref. \cite{2006JCoPh.217..834M} in Eqs.~(9) and (10) of his work. By evaluating integrals of the form $\int_{\lambda_{\min}}^{\lambda_{\max}} \frac{1}{\omega^2+\lambda^2} \frac{d\lambda}{\lambda^\beta}$, it is demonstrated that a continuous superposition of single exponential relaxation processes precisely generates an intermediate power-law spectral region. In a strikingly similar fashion, our kinetic model naturally produces the fundamental response spectra of the plasma (e.g., Lorentzian-like modes or highly damped transients). The superposition of these basis spectra, weighted by the power-law distribution of the stochastic correlation times $\lambda$, is precisely the mechanism that intrinsically generates these $1/f^\alpha$ macroscopic noise signatures.

\subsection{Link with in-situ observations}
The closed-form expressions derived above, such as Eqs.~\eqref{closedexpr1} and \eqref{closedexpr2}, can be employed for assessing whether in-situ observations match with the derived theoretical expressions, thus giving the possibility to interpret space plasmas within the developed FDR framework. Indeed, analytical solutions can serve as theoretical functions to fit in-situ measurements at high frequency, bypassing the standard practice of piecewise linear fitting on log-log scales. Assuming the validity of the Taylor hypothesis to map the observed frequencies to spatial wavenumbers, a model-based fitting procedure, as performed in Refs. \cite{Carbone_2022, Chiappetta_2023}, allows to extract the forcing parameter such as the exponent $\mu$ and the memory cutoffs $\lambda_{\min}$ and $\lambda_{\max}$ from in-situ measurements.

Before transitioning to the observable quantities, it is however important to point out that if a flat spatial spectrum is assumed for the stochastic forcing, i.e., $|q_k|^2 = \text{const}$, the model predicts non-physical features for the electric field fluctuations, whose spectrum is given by $\langle|E_k|^2\rangle = k^2 \langle|\varphi_k|^2\rangle$. Specifically, multiplying the asymptotic potential scalings by $k^2$ would lead in some cases to flat or even growing electric field spectra at sub-ion scales, heavily contradicting standard spacecraft observations in the solar wind. This apparent shortcoming arises because a flat $|q_k|^2$ implies that turbulent energy is artificially injected with equal intensity at all spatial scales, thus bypassing the inertial cascade process. A natural solution to this issue is to assume that the power injection from the larger scales, modeled by the correlated noise, is itself the result of an ongoing turbulent cascade. Therefore, it is physically sound to characterize the spatial profile of the forcing with a spectral scaling of the form $|q_k|^2 \propto k^{-\delta}$, where $\delta$ represents the spectral index of the incoming inertial cascade. Under this assumption, the collective plasma response described by Vlasov theory acts effectively as a low-pass filter over the background turbulence. Table~\ref{tab:electric_field_scalings} summarizes the asymptotic spectral scalings of the electric field, derived from the single-$\lambda$ FDR \eqref{FDRsinglelambda} in Sec.~\ref{sec2.1} by assuming a spatially correlated noise source with spectral index $\delta$.

\begin{table}[htbp]
    \centering
    \renewcommand{\arraystretch}{1.5}
    \begin{tabular}{|l|l|c|}
        \hline
        \textbf{Plasma Regime} & \textbf{Correlation Regime} & \textbf{ Scaling } \\
        \hline
        
        \multirow{2}{*}{$\alpha \to -1$} 
        & High correlation ($\lambda \ll \gamma_L$) & \qquad  \makebox[1cm][l]{$k^{-\delta}$} \\ \cline{2-3}
        & Low correlation ($\lambda \gg \gamma_L$) & \qquad  \makebox[1cm][l]{$k^{1-\delta}$} \\
        \hline
        
        \multirow{2}{*}{$\alpha \to \infty$} 
        & High correlation ($\lambda \ll \sqrt{\alpha}k$) & \qquad \makebox[1cm][l]{$k^{-1-\delta}$} \\ \cline{2-3}
        & Low correlation ($\lambda \gg \sqrt{\alpha}k$) & \qquad \makebox[1cm][l]{$k^{1-\delta}$} \\
        \hline
        
        $k \to \infty$ & Independent of $\lambda$ & \qquad 
        \makebox[1cm][l]{$k^{-\delta}$} \\
        \hline
    \end{tabular}
    \caption{Asymptotic spectral scalings for the electric field fluctuations $\langle|E_k|^2\rangle$, assuming an injection spectrum for the spatial forcing $|q_k|^2 \propto k^{-\delta}$ for different plasma ($\alpha$) and correlation regimes. The low correlation case corresponds to the white-noise forcing.}
    \label{tab:electric_field_scalings}
\end{table}

The strategy employed for the single-$\lambda$ case can be straightforwardly extended to the scenario where the correlation rate $\lambda$ is governed by a power-law distribution. By applying the exact same procedure, i.e multiplying the electrostatic potential spectrum by $k^2$ and incorporating the inertial cascade spectrum $|q_k|^2 \propto k^{-\delta}$, one can readily deduce the corresponding electric field scalings from the $\lambda$-averaged asymptotic limits derived previously. More generally, this framework is not restricted to asymptotic approximations. It can be applied directly to the full, non-asymptotic kinetic FDR averaged over the $\lambda$ distribution given in Eq.~\eqref{FDRaverage}, thereby providing a complete and exact theoretical prediction for the electric field spectrum $\langle|E_k|^2\rangle$ across all relevant spatial scales. 

To conclude this section we present the asymptotic behaviors of the electric field spectra in two specific cases.
First, we consider  the $\alpha\to -1$ case in which the characteristic timescale of the system, i.e. the Landau damping rate $\gamma_L$, vanishes in the limit. Therefore, for a fixed range $[\lambda_{min},\lambda_{max}]$, the low-correlation regime $\lambda\gg \gamma_L$ is, at some point, inevitable and much more likely to occur.
Second, we consider the the $\alpha\to\infty$ limit in which the Langmuir frequency $\omega_L=\sqrt{\alpha}k$ becomes infinitely large and the high-correlation regime $\lambda\ll\omega_L$ is the most likely configuration.

\begin{figure}[hbtp!]
\centering
\includegraphics[width=8cm]{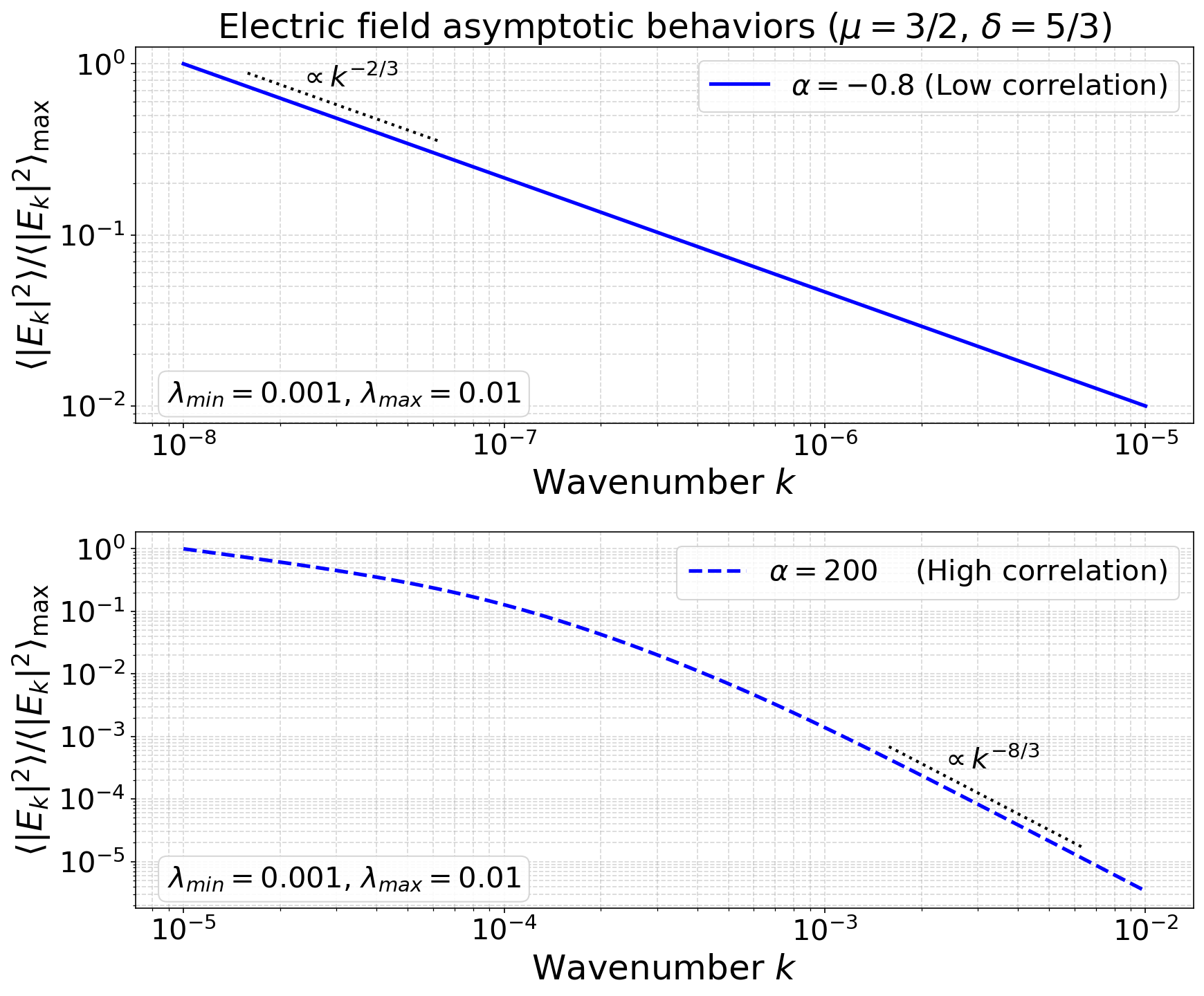}
\caption{Asymptotic behavior of the electric field for a correlated noise spectral index $\delta=5/3$. \textit{Upper panel:} the electric field spectral index  in the $\alpha\to-1$ limit low correlation regime is $-2/3$.  \textit{Lower panel:} the electric field spectral index  in the $\alpha\to\infty$ limit high correlation regime is $-8/3$.}
\label{fig:elecspectralbreak}
\end{figure}

In Fig.~\ref{fig:elecspectralbreak} we display the electric field spectra obtained from Eq.~\eqref{FDRaverage} in $i)$ for $\alpha=-0.8$ and $k\in[10^{-8},10^{-5}]$ and in $ii)$ for $\alpha=200$ and $k\in[10^{-5},10^{-2}]$. In both cases we set $\mu=3/2$, $\lambda_{min}=10^{-3}$, $\lambda_{max}=10^{-2}$, so that $i)$ and $ii)$ are representative of low and high correlation regimes, respectively. We fix the noise spectral index $\delta=5/3$, to model a Kolmogorov-like behavior for the energy cascade at large (fluid) scales.

Under these assumptions the model predicts an asymptotic electric field spectral index equal to $2/3$ in $i)$ and $8/3$ in $ii)$ as it can be easily appreciated from Fig.~\ref{fig:elecspectralbreak}. We outline that these spectral slopes are qualitatively in agreement with recent observations in the turbulent Earth's magnetosheath\cite{48ys-3m6m} in the ionic and electronic ranges, respectively. However, this correspondence should be taken with caution. Indeed, our model is very simplified as it focuses on electrostatic fluctuations and it is one-dimensional, and it does not explicitly include any microphysics plasma processes as the peculiar type of waves that may be present in the system. Further studies are necessary to better comprehend if this qualitative agreement is robust.

\section{Conclusions}\label{sec5}
In this work, we have developed a generalized kinetic framework to model plasma turbulence driven by a stochastic source with a finite correlation time. By constructing a kinetic analogue of the Langevin equation, we derived an exact FDR that characterizes the steady-state balance between the energy injected by the correlated noise and the collisionless dissipation provided by Landau damping. This framework allows for the exploration of diverse physical scenarios, ranging from electronic Langmuir waves to slow-MHD compressive modes, by tuning the coupling parameter $\alpha$.  

We analytically extracted the asymptotic spectral scalings of the electrostatic potential in several limits of interest, including the weakly damped non-oscillatory limit ($\alpha \to -1$) and the weakly damped fast-wave limit ($\alpha \to \infty$). Furthermore, the investigation of the $k \to \infty$ limit revealed that at sub-ion and sub-electron scales, kinetic dynamical effects are entirely suppressed, and the system responds purely through static screening. The turbulent cascade in velocity space was also formally characterized via a Hermite decomposition, confirming an $m^{-1/2}$ asymptotic behavior. All these analytical predictions successfully reproduce the white-noise phenomenology \cite{2015JPlPh..81a3004K} when the vanishing correlation time limit $\lambda\to\infty$ is considered. Subsequently, we validated the asymptotic spectral scalings and the Hermite asymptotic spectrum through direct numerical integrations of the underlying collisionless Vlasov system.  

To bridge the gap between theoretical modeling and in-situ spacecraft observations, we extended our analysis to a multiscale forcing scenario where the noise correlation rate $\lambda$ is treated as a stochastic variable governed by a power-law distribution. This extension mathematically mirrors the superposition principles that generate $1/f^\alpha$ macroscopic noise signatures. By averaging the single-$\lambda$ FDR over this distribution, we derived exact, closed-form expressions for the spectra. These analytical solutions are intended to serve as a library of theoretical fitting functions for observational data in the solar wind and magnetosphere, circumventing the limitations of standard piecewise linear fitting.  

Finally, to ensure the physical consistency of the predicted electric field fluctuations at small scales, we incorporated an inertial cascade spectral index into the spatial forcing profile, assuming $|q_k|^2 \propto k^{-\delta}$. Under this hypothesis, the Vlasov plasma effectively acts as a kinetic low-pass filter over the background turbulence. This framework could provide a  model-based method, agnostic with respect to the underlying kinetic plasma process at play, to interpret electric field spectrum at kinetic scales observed in space.  

\section*{Author Declarations}
\subsection*{Data availability statement}
\noindent The data that supports the findings of this study are available from the corresponding author upon reasonable request.
\subsection*{Conflict of interest statement}
\noindent The authors have no conflicts to disclose.
\begin{acknowledgments}
FM, OP, and AL are supported by the FIS2 Starting Grant FIS-2023-00246 “PhAse-sPAce cOmplexity in turbulent nearly-reversible plasmas (PAPAO)” (CUP B53C24009610001) funded by the Italian Ministry of University and Research. 
SB and GC acknowledge support from the Space It Up! project funded by the Italian Space Agency, ASI, and the Ministry of University and Research, MUR, under Contract
No. 2024-5-E.0-CUP n. I53D24000060005.
%SB, GC and LSV were supported by the Agenzia Spaziale Italiana (ASI) project ``Radial Evolution of large- and KInetic-scale Processes in the Expanding Solar wind (REKIPES)'' C83C25000900005.
LSV was supported by the Swedish National Space Agency (SNSA) Research Grant N. 2025-00229.
\end{acknowledgments}

\bibliographystyle{unsrt}
\bibliography{biblio}

\end{document}